
\output={\plainoutput}

\newcount\pagenumber
\newcount\questionnumber
\newcount\sectionnumber
\newcount\appendixnumber
\newcount\equationnumber
\newcount\referencenumber
\newcount\subsecnumber

\global\subsecnumber=1

\def\ifundefined#1{\expandafter\ifx\csname#1\endcsname\relax}
\def\docref#1{\ifundefined{#1} {\bf ?.?}\message{#1 not yet defined,}
\else \csname#1\endcsname \fi}

\newread\bib
\newwrite\reffs

\newcount\linecount
\newcount\citecount
\newcount\localauthorcount

\def\article{
\def\eqlabel##1{\edef##1{\sectionlabel.\the\equationnumber}}
\def\seclabel##1{\edef##1{\sectionlabel}}
\def\feqlabel##1{\ifnum\passcount=1
\immediate\write\crossrefsout{\relax}  
\immediate\write\crossrefsout{\def\string##1{\sectionlabel.
\the\equationnumber}}\else \fi }
\def\fseclabel##1{\ifnum\passcount=1
\immediate\write\crossrefsout{\relax}   
\immediate\write\crossrefsout{\def\string##1{\sectionlabel}}\else\fi}
\def\cite##1{\immediate\openin\bib=bib.tex\global\citecount=##1
\global\linecount=0{\loop\ifnum\linecount<\citecount \read\bib
to\temp \global\advance\linecount by 1\repeat\temp}\immediate\closein\bib}
\def\docite##1 auth ##2 title ##3 jour ##4 vol ##5 pages ##6 year ##7{
\par\noindent\item{\bf\the\referencenumber .}
 ##2, ##3, ##4, {\bf ##5}, ##6,
(##7).\par\vskip-0.8\baselineskip\noindent{
\global\advance\referencenumber by1}}
\def\dobkcite##1 auth ##2 title ##3 publisher ##4 year ##5{
\par\noindent\item{\bf\the\referencenumber .}
 ##2, {\it ##3}, ##4, (##5).
\par\vskip-0.8\baselineskip\noindent{\global\advance\referencenumber by1}}
\def\doconfcite##1 auth ##2 title ##3 conftitle ##4 editor ##5 publisher ##6
year ##7{
\par\noindent\item{\bf\the\referencenumber .}
##2, {\it ##3}, ##4,  {edited by: ##5}, ##6, (##7).
\par\vskip-0.8\baselineskip\noindent{\global\advance\referencenumber by1}}}

\def\normalarticlestyle{
\immediate\openout\reffs=reffs
\global\referencenumber=1
\def\eqlabel##1{\edef##1{\sectionlabel.\the\equationnumber}}
\def\seclabel##1{\edef##1{\sectionlabel}}
\def\feqlabel##1{\ifnum\passcount=1
\immediate\write\crossrefsout{\relax}  
\immediate\write\crossrefsout{\def\string##1{\sectionlabel.
\the\equationnumber}}\else \fi }
\def\fseclabel##1{\ifnum\passcount=1
\immediate\write\crossrefsout{\relax}   
\immediate\write\crossrefsout{\def\string##1{\sectionlabel}}\else\fi}
\def\z@{ 0pt}
\def\cite##1{\immediate\openin\bib=bib.tex\global\citecount=##1
\global\linecount=0{\loop\ifnum\linecount<\citecount \read\bib
to\temp \global\advance\linecount by
1\repeat\immediate\write\reffs{\temp}\global\advance\referencenumber by1}
\immediate\closein\bib}
\def\docite##1 auth ##2 title ##3 jour ##4 vol ##5 pages ##6 year ##7{
\par\noindent\item{\bf\the\referencenumber .}
 ##2, ##3, ##4, {\bf ##5}, ##6,
(##7).\par\vskip-0.8\baselineskip\noindent}
\def\dobkcite##1 auth ##2 title ##3 publisher ##4 year ##5{
\par\noindent\item{\bf\the\referencenumber .}
 ##2, {\it ##3}, ##4, (##5).
\par\vskip-0.8\baselineskip\noindent}
\def\doconfcite##1 auth ##2 title ##3 conftitle ##4 editor ##5 publisher ##6
year ##7{
\par\noindent\item{\bf\the\referencenumber .}
##2, {\it ##3}, ##4,  {edited by: ##5}, ##6, (##7).
\par\vskip-0.8\baselineskip\noindent}}

\def\appendixlabel{\ifcase\appendixnumber\or A\or B\or C\or D\or E\or
F\or G\or H\or I\or J\or K\or L\or M\or N\or O\or P\or Q\or R\or S\or
T\or U\or V\or W\or X\or Y\or Z\fi}

\def\sectionlabel{\ifnum\appendixnumber>0 \appendixlabel
\else\the\sectionnumber\fi}

\def\beginsection #1
 {{\global\appendixnumber=0\global\advance\sectionnumber by1}\equationnumber=1
\par\vskip 0.8\baselineskip plus 0.8\baselineskip
 minus 0.8\baselineskip
\noindent$\S$ {\bf \the\sectionnumber . #1}
\par\penalty 10000\vskip 0.6\baselineskip plus 0.8\baselineskip
minus 0.6\baselineskip \noindent}

\def\subsec #1 {\bf\par\vskip8truept  minus 8truept
\noindent \ifnum\appendixnumber=0 $\S\S\;$\else\fi
$\bf\sectionlabel.\the\subsecnumber$ #1
\global\advance\subsecnumber by1
\rm\par\penalty 10000\vskip6truept  minus 6truept\noindent}

\def\beginappendix #1
{{\global\advance\appendixnumber by1}\equationnumber=1\par
\vskip 0.8\baselineskip plus 0.8\baselineskip
 minus 0.8\baselineskip
\noindent
{\bf Appendix \appendixlabel . #1}
\par\vskip 0.8\baselineskip plus 0.8\baselineskip
 minus 0.8\baselineskip
\noindent}

\def\no{\eqno({\rm\sectionlabel}
.\the\equationnumber){\global\advance\equationnumber by1}}

\def\beginref #1 {\par\vskip 2.4 pt\noindent\item{\bf\the\referencenumber .}
\noindent #1\par\vskip 2.4 pt\noindent{\global\advance\referencenumber by1}}

\def\ref #1{{\bf [#1]}}
\magnification=1200
\output={\plainoutput}

\normalarticlestyle
\advance\hsize by10truept
\advance\vsize by10truept
\headline={{\rm\ifnum\pageno=1\hfill\else\ifodd\pageno
Nash and O' Connor\hfill
BRST Quantisation and the Product
Formula...\else BRST Quantisation and the
Product Formula...\hfill Nash and O' Connor\fi\fi}}
\def\det{Det\,}
\def\ee{{\cal E}}
\def\mapright#1{\smash{\mathop{\longrightarrow}\limits^{#1}}}
\def\mapleft#1{\smash{\mathop{\longleftarrow}\limits^{#1}}}
\def\ker{ker\,}
\def\im{Im\,}
\def\Det{{\rm Det}\,}

\def\r{\rightarrow}
\def\ref#1{{\bf[#1]}}
\def\e{{\varepsilon}}
\def\f{{\bar f}}
\def\tfb{{\kern 3pt{\widetilde{\kern -3pt\f}}\kern -1pt}}
{\hfill dias-stp-93-22}
\par\vfill
\centerline{BRST Quantisation and the Product
Formula for the Ray-Singer Torsion}
\vskip1.25\baselineskip
\centerline{by}
\vskip1.25\baselineskip
\centerline{Charles Nash$^{*,{\dag}}$ and Denjoe O' Connor$^{\dag}$}
\par\vskip\baselineskip
\noindent
$^*$Department of Mathematical Physics\hfill
$^{\dag}$School of Theoretical Physics
\par\noindent
St. Patrick's College\hfill              Dublin Institute for Advanced
Studies
\par\noindent
Maynooth\hfill                {\it and}\hfill  10 Burlington Road
\par\noindent
Ireland\hfill                            Dublin 4
\par\noindent
\null\hfill                              Ireland
\par\vskip3\baselineskip
\noindent
{{\bf Abstract}: We give a quantum field theoretic derivation of the
formula obeyed by the Ray-Singer torsion on product manifolds. Such
a derivation has proved elusive up to now. We use a BRST formalism
which introduces the idea of an infinite dimensional Universal Gauge
Fermion, and is of independent interest being applicable to
situations other than the ones considered here. We are led to a new
class of Fermionic topological field theories. Our methods are also
applicable to combinatorially defined manifolds and methods of
discrete approximation such as the use of a simplicial lattice or
finite elements. The topological field theories discussed provide a
natural link between the combinatorial and analytic torsion.}
\beginsection{Introduction}
Quantum field theories have, since their inception, received a
considerable onslaught from a variety of mathematical techniques,
particularly those drawn from analysis. However in the last decade
there has been ample evidence that the way forward is considerably
illuminated if techniques from topology are used. It has been
learned, particularly in the case of gauge theories, that quantum
and statistical fluctuations are sensitive to the global properties
of the manifold on which they occur, topological field theories
being, perhaps,  the ultimate example.
\par
In this paper we discuss the field theoretic realizations of the
Ray-Singer torsion, which arises as a power of the partition
function of an appropriate topological field theory. This subject
has of course been considered before however one obstacle remained,
and that was a field theoretic demonstration of the product formula
(cf. Birmingham et al. \ref{1}) which states that the torsion on a
product manifold, $M=M_1\times M_2$, is the torsion of one factor
raised to the the Euler character of the the other factor. The
principal purpose of this paper is therefore to give a field
theoretic demonstration of this formula.
\par
The techniques we employ are a mixture of harmonic analysis and
field theory. A detailed understanding of the harmonic analysis is
necessary in order to establish certain isomorphisms, and to
construct a  detailed decomposition of the harmonic forms
appropriate for a product manifold. The principal results of the
paper are
\item{(i)}The decomposition of the kernels of $d$ and $d^*$ on a
product manifold in terms of appropriate data on the factors.
\item{(ii)}The construction of the gauge fixing sector for an
$n$-form field in arbitrary dimensions, together with the
decomposition of this on a product manifold.
\item{(iii)}The introduction of a new class of Fermionic topological
field theories, and a larger graded class of topological field
theories whose partition function $Z$ is the quarter power of the
torsion $T$ ($Z=T^{1/4}$), for all $M$. This latter graded class
exists on all odd dimensional manifolds $M$; and if  $\dim M=2n+1$,
then the corresponding topological field theories are {\it Bosonic}
for $n$ {\it odd}  and {\it Fermionic} for $n$ {\it even}.
\item{(iv)}A proof of the product formula in topological field
theory. This field theory approach makes manifest the connection
between the analytic and combinatorial torsions since the
topological BRST action that arises is easily considered on a
simplicial lattice. This amounts to one of many possible
regularisation procedures in field theory. The convergence of the
eigenvalues of the combinatorial Laplacian to those of the continuum
Laplacian (see M\"uller \ref{2}) then establishes the equivalence of
the combinatorial and analytic torsions in the field theory context.
\par\noindent
The layout of the paper is as follows. In section 2 we introduce the
Ray-Singer torsion, and analyse of the Laplacians on $r$-forms,
establish some isomorphisms between the kernel of $d$ and the kernel
of $d^*$ for these spaces and the associated ``shifted Poincar\'e
duality''. We establish the corresponding decomposition for product
manifolds. The section ends with a simple derivation  of the product
formula for the torsion. The starting point for this analysis is the
definition of the analytic Ray-Singer torsion in terms of zeta
functions.
\par
Section 3 contains a description of the BRST method for constructing
the path integral measure for the quantisation of an $n$-form gauge
field where the action for this field is invariant under the
addition to this field of a closed $(n-1)$-form field. The method is
the gauge Fermion construction of Batalin and Vilkovisky \ref{7,8}.
We introduce a powerful notation for handling this construction,
whereby the dimension of the manifold plays a minor r\^ole.  The
gauge Fermion triangle is formally infinite, the gauge fixing
necessary is carried out by a certain finite sub-triangle
determined by  the particular field to be found at its apex. We
finish the section by considering the gauge Fermion on product
manifolds. The procedures used are purely algebraic and do not rely
on the manifold being a differentiable one. The results are readily
translatable to a simplicial manifold.
\par
In section 4 we introduce our topological field theories. By
choosing the field which forms the lower apex of the gauge Fermion
triangle to be Fermionic we find an interesting class of topological
field theories, which alternates between Fermionic and Bosonic as
the dimension changes. Each theory has the torsion to the quarter
power as its partition function. We also consider briefly the wider
class of topological actions where two fields are used  and show
that these have the torsion to an appropriate power as partition
function. We finish the section by considering these theories on
product manifolds. An analysis of the resulting  partition function
readily gives a derivation of the product formula. This procedure is
readily implementable on simplicial manifolds and can be interpreted
as establishing the equivalence of the combinatorial and analytic
torsions.
\par
\beginsection{The Ray-Singer torsion} The Ray-Singer torsion, which
we denote by $T(M,E)$, is a positive real number which is defined
when one has a flat connection $A$ on a bundle $E$ over a compact
manifold $M$ together with a certain representation of the
fundamental group $\pi_1(M)$.
\par
In brief the definition of $T(M,E)$ arises as follows: For a flat
connection $A$ on $M$ to give rise to  a non-trivial situation one
requires $A$ to have non-zero holonomy  and thus $M$ must have a
non-trivial $\pi_1(M)$. The holonomy of $A$ round a based loop $C$
is then realised in the usual way by the path ordered integral
$$h(C)=P\exp\left[\int_CA\right]\no$$
These $h(C)$ then provide a representation  of $\pi_1(M)$, the one
referred to above, in the gauge group $G$ of $A$. For the torsion
$T(M,E)$ one must consider the volume elements $V^r$ associated with
each cohomology group $H^r(M;E)$. Determinants give a natural
definition for volume elements and, in terms of de Rham cohomology,
one can realise these $V^r$ as determinants of Laplacians
$$\det \Delta^E_r\no$$
where
$$ \Delta^E_r=\left(d_A d^*_A+d^*_A d_A\right)\no$$
is the Laplacian on $r$-forms with coefficients in $E$ and $d_A^2=0$
since $A$ is flat\footnote{$^*$}{Because of this fact, from now on,
we shall abbreviate $d_A$ to simply $d$.}. Zeta functions can be
used to define these determinants provided $\Delta^E_r$ is positive
which is assumed. The torsion is then defined by
\eqlabel{\tordef}
$$\eqalign{\ln T(M,E)&=\sum_{r=0}^n(-1)^rr\ln \det\Delta^E_r,
                                                       \qquad n=\dim M\cr
 &=-\sum_{r=0}^n(-1)^rr\zeta^\prime_{\Delta^E_r}(0)\cr}\no$$
where we have used the standard definition of a determinant via the
zeta function: i.e. if $O$ is an elliptic differential or
pseudo-differential operator with positive spectrum $\{\lambda_n\}$
then $\det O$ is defined via
$$\ln\det O=-\zeta^\prime_O(0),\qquad\hbox{where }
\zeta_O(s)=\sum_{\lambda_n}{\Gamma_n\over\lambda_n^s}\no$$
and $\Gamma_n$ denotes the degeneracies of the $\lambda_n$.
For convenience we also write
\eqlabel{\taudef}
$$\eqalign{\ln T(M,E)&=\left.{d \tau(s)\over ds}\right\vert_{s=0}\cr
\hbox{where }\qquad \tau(s)&=-\sum_{r=0}^n(-1)^rr\zeta_{\Delta^E_r}(s)\cr}\no$$
\par
Some standard properties \ref{2,3,4} of $T(M,E)$
are:\footnote{${}^{\dag}$}{$M$ is always a compact closed Riemannian
manifold}
\item{(i)}   $T(M,E)>0$
\item{(ii)}  $T(M,E)=1$ if $\dim M$ is even
\item{(iii)} $T(M,E)$ is {\it independent} of the metric used to
construct the $\Delta^E_r$.
\par
Property (ii) above means that the non-trivial cases occur, in the main,
when $M$ is odd-dimensional; although one can sometimes make
non-trivial deductions in which $T(M,E)$ plays a r\^ole
when $M$ is even dimensional e.g. if $M$ is a complex manifold
cf. Ray-Singer \ref{5} and Witten \ref{6}.
Property (iii) asserts that $T(M,E)$ is a metric independent quantity
and suggests, what is in fact true, that $T(M,E)$ may be a topological
invariant of $M$.
\par
We now need to establish some notation and also properties of the
torsion relevant for us here. Central to the discussion are the spectra
of the Laplacians $\Delta^E_r$; so we define $\ee_r(M,\lambda)$ to be
the eigenspace of $\Delta^E_r$ corresponding to the eigenvalue
$\lambda$, i.e.
$$\ee_r(M,\lambda)=\{\omega\in\Gamma(M,\Lambda^rT^*M\otimes E):\,
\Delta^E_r\omega=\lambda\omega\}\no$$
where $\Gamma(M,\Lambda^rT^*M\otimes E)$ is the space of sections, over
$M$,  of the $r$-form bundle $\Lambda^rT^*M\otimes E$. Now this eigenspace
$\ee_r(M,\lambda)$ has a decomposition \ref{2} which is particularly useful for
the subsequent quantum field theory. Recall that $d$ and $d^*$ act  as shown
below
$$\eqalign{&\cdots\mapright{d}
\ee_r(M,\lambda)\mapright{d}\ee_{r+1}(M,\lambda)\mapright{d}\cdots\cr
&\cr
&\cdots\mapleft{d^*}
\ee_r(M,\lambda)\mapleft{d^*}\ee_{r+1}(M,\lambda)\mapleft{d^*}\cdots\cr}\no$$
and so
$$\ee_r(M,\lambda)\simeq\im d\oplus\ker d^*\simeq\im d^*\oplus\ker d\no$$
But our positivity requirement means that
$$\ker d\cap\ker d^*=\phi$$
and it is then easy to check that if we further define
$$\ee_r^\prime(M,\lambda)=\{\omega\in\ee_r(M,\lambda):\,d\omega=0\}\qquad
 \ee_r^{\prime\prime}(M,\lambda)=\{\omega\in\ee_r(M,\lambda):\,d^*\omega=0\}$$
then $\ee_r(M,\lambda)$ has the decomposition
\eqlabel{\decomp}
$$\ee_r(M,\lambda)=\ee_r^\prime(M,\lambda)
\oplus\ee_r^{\prime\prime}(M,\lambda)\no$$
allowing one to verify that
\eqlabel{\isomor}
$$\ee_r^{\prime\prime}(M,\lambda)\mapright{d/\sqrt{\lambda}}
\ee^{\prime}_{r+1}(M,\lambda)\no$$
is an isomorphism with inverse $d^*/\sqrt{\lambda}$ so that
\eqlabel{\isomor}
$$\ee_r^{\prime\prime}(M,\lambda)\simeq \ee^{\prime}_{r+1}(M,\lambda)\no$$
\par
For the computation of the torsion it is useful to display the
information encapsulated in \docref{decomp} and \docref{isomor} above in a
table. Doing this we have
\eqlabel{\table}
$$\eqalign{\ee_n(M,\lambda)&=\ee_n^\prime\oplus\phi\cr
                           &\quad\qquad\nwarrow\cr
    \ee_{n-1}(M,\lambda)&=\ee_{n-1}^{\prime}\oplus\ee_{n-1}^{\prime\prime}\cr
&\vdots\cr
\ee_{r}(M,\lambda)&=\ee_{r}^{\prime}\oplus\ee_{r}^{\prime\prime}\cr
&\quad\qquad\nwarrow\cr
\ee_{r-1}(M,\lambda)&=\ee_{r-1}^{\prime}\oplus\ee_{r-1}^{\prime\prime}\cr
&\vdots\cr
\ee_{1}(M,\lambda)&=\ee_{1}^{\prime}\oplus\ee_{1}^{\prime\prime}\cr
&\qquad\nwarrow\cr
\ee_0(M,\lambda)&=\phi\oplus\ee_0^{\prime\prime}\cr}
\no$$
Note that the arrow in \docref{table} simply connects pairs of spaces which
are isomorphic.
\par
A final stage in making use of the refinement of the
spectral data embodied in the decomposition table above is to restrict
the Laplacians $\Delta_r^E$ to the spaces $\ee_r^\prime(M,\lambda)$ and
$\ee_r^{\prime\prime}(M,\lambda)$. These two restrictions correspond
precisely to the operators $dd^*$ (acting on $\ee_r^\prime(M,\lambda)$) and
$d^*d$ (acting on $\ee_r^{\prime\prime}(M,\lambda)$) respectively.
In this way we construct the associated zeta functions for these
operators giving
$$\eqalign{\zeta_{dd^*_r}(s)&
=\sum_{\lambda}{\Gamma_r^\prime(\lambda)\over\lambda^s},\qquad
\zeta_{d^*d_r}(s)=\sum_{\lambda}{\Gamma_r^{\prime\prime}(\lambda)
\over\lambda^s}\cr
\hbox{and }\qquad \zeta_{\Delta_r^E}(s)
&=\zeta_{dd^*_r}(s)+\zeta_{d^*d_r}(s)\cr}\no$$
in an obvious notation, and clearly
$\Gamma_r(\lambda)=\Gamma_r^\prime(\lambda)+\Gamma_r^{\prime\prime}(\lambda)$.
However if we now return to the definition \docref{tordef} of the
torsion $T(M,E)$ which was
$$\ln T(M,E)=-\sum_0^{n}(-1)^rr\zeta^\prime_{({dd^*+d^*d})_r}(0)\no$$
and use the information contained in table \docref{table}, we find
that
$$\ln T(M,E)=-\sum_0^{n-1}(-1)^r\zeta^\prime_{{d^*d}_r}(0)\no$$
Of course an analogous formula exists for $\tau(s)$.
\par
It will also be useful later to take account of isomorphisms that
arise because of Poincar\'e duality. In the first instance
Poincar\'e duality asserts that
$$\ee_r(M,\lambda)\simeq\ee_{n-r}(M,\lambda)\no$$
Refining this with respect to the decomposition \docref{decomp}
gives the further statement that
$$\eqalign{\ee_r^\prime(M,\lambda)
&\simeq\ee_{n-r}^{\prime\prime}(M,\lambda)\cr
  \ee_r^{\prime\prime}(M,\lambda)&\simeq\ee_{n-r}^{\prime}(M,\lambda)\cr}\no$$
Finally if we use our table above we find that a \lq shifted
Poincar\'e duality' exists for the spaces
$\ee_r^{\prime\prime}(M,\lambda)$ alone, namely
\eqlabel{\poincare}
$$\ee_r^{\prime\prime}(M,\lambda)
\simeq\ee_{n-r-1}^{\prime\prime}(M,\lambda)\no$$
This gives a further refinement to the formula for the torsion since we
can now easily verify that, for {\it odd } $n$, $n=2m+1$, one has
\eqlabel{\shiftedtorformula}
$$\ln T(M,E)
=-2\sum_{r=0}^{m-1}(-1)^r\zeta^\prime_{{d^*d}_r}(0)
-{(-1)}^m\zeta^\prime_{d^*d}(0),\qquad\hbox{where }n=2m+1\no$$
and for {\it even} $n$, $n=2m$ we have a cancellation of terms
rather than a doubling and hence
$$\ln T(M,E)=0$$
\par
This completes our brief account of the properties of the torsion
that we need for this paper. We are particularly interested in
$T(M,E)$ for the case where
$$M=M_1\times M_2\no$$
and we turn to this matter in the next section.
\subsec{The Decomposition of $\ee^\prime_r$ and $\ee^{\prime\prime}_{r}$ on
Product Manifolds}
We shall now assume that
$$M=M_1\times M_2,\qquad \hbox{with } \dim M \hbox{ odd}\no$$
we also take the convention that $\dim M_1$ is odd and $\dim M_2$ is
even. In this situation if we take $M_2$  to be simply connected, so
that it supports no non-trivial flat connections, then as we shall
demonstrate below $T(M,E)$ has the property that
$$T(M,E)=\left\{T(M,E_1)\right\}^{\chi(M_2)}\no$$
where $E_1$ denotes the flat bundle over $M_1$ and $\chi(M_2)$ is the
usual Euler-Poincar\'e characteristic of $M_2$.
\par
Now using the  notation defined above on $M_1\times M_2$ we have
\eqlabel{\torformula}
$$\eqalign{\ee_r(M_1\times M_2,\nu)&=\ee_r^\prime(M_1\times M_2,\nu)
\oplus\ee_r^{\prime\prime}(M_1\times M_2,\nu)\cr
&={\textstyle\bigoplus\limits_{p+q=r}}\ee_p(M_1,\lambda)
\wedge\ee_q(M_2,\mu),\qquad\hbox{with }\nu=\lambda+\mu\cr}\no$$
Vital for us is the relation implicit in eq. \docref{torformula}
above between the spaces on its RHS.  In particular we need an
explicit construction of the space $\ee_r^{\prime\prime}(M_1\times
M_2,\nu)$. This we now provide. The appropriate projection operator
is $d^*d$ and so, formally, we have
$$\ee_r^{\prime\prime}(M_1\times M_2,\nu)
={d d^*\over\nu}\ee_r(M_1\times M_2,\nu)\no$$
Hence we write
$$\eqalign{\ee_r^{\prime\prime}(M_1\times M_2,\nu)&={\textstyle
\bigoplus\limits_{p+q=r}}
{d^*d\over\nu}\left(\ee_p(M_1,\lambda)\wedge\ee_q(M_2,\mu)\right)\cr
&={\textstyle
\bigoplus\limits_{p+q=r}}{d^*\over\nu}\left\{
d\ee_p(M_1,\lambda)
\wedge \ee_q(M_2,\mu)\oplus (-1)^p\ee_p(M_1,\lambda)\wedge
d\ee_q(M_2,\mu)\right\}\cr
&={\textstyle
\bigoplus\limits_{p+q=r}}{d^*\over\nu}\left\{
d\ee_p^{\prime\prime}(M_1,\lambda)
\wedge \ee_q(M_2,\mu)\oplus (-1)^p\ee_p(M_1,\lambda)\wedge
d\ee_q^{\prime\prime}(M_2,\mu)\right\}\cr
&={\textstyle
\bigoplus\limits_{p+q=r}}{1\over\nu}\left\{
d^*d\ee_p^{\prime\prime}(M_1,\lambda)
\wedge \ee_q(M_2,\mu)\oplus(-1)^{p+1}
d\ee_p^{\prime\prime}(M_1,\lambda)
\wedge d^*\ee^\prime_q(M_2,\mu)
\right.\cr
&\oplus\left.(-1)^p d^*\ee_p^{\prime}(M_1,\lambda)
\wedge d\ee^{\prime\prime}_q(M_2,\mu)\oplus(-1)^{2p}
\ee_p(M_1,\lambda)\wedge d^*d\ee^{\prime\prime}_q(M_2,\mu)\right\}
\cr}
\no$$
\par
Next we must use the information from table \docref{table}
and, doing this, we find that
$$\eqalign{\ee_r^{\prime\prime}(M_1\times M_2,\nu)
&={\textstyle
\bigoplus\limits_{p+q=r}}{1\over\nu}\left\{
\lambda\ee_p^{\prime\prime}(M_1,\lambda)
\wedge \ee_q(M_2,\mu)\right.\cr
&\left.\oplus(-1)^{p+1}
\sqrt{\lambda\mu}\ee_{p+1}^{\prime}(M_1,\lambda)
\wedge \ee^{\prime\prime}_{q-1}(M_2,\mu)
\right.\cr
&\oplus\left.(-1)^p
\sqrt{\lambda\mu}\ee_{p-1}^{\prime\prime}(M_1,\lambda)
\wedge \ee^{\prime}_{q+1}(M_2,\mu)\oplus\mu
\ee_p(M_1,\lambda)\wedge \ee^{\prime\prime}_q(M_2,\mu)\right\}
\cr}
\no$$
Expanding now the terms $\ee_q(M_2,\mu)$ and $\ee_p(M_1,\lambda)$,
these being the only  terms in the above without a prime or a
double-prime, we get the (temporarily) longer expression
$$
\eqalign{\ee_r^{\prime\prime}(M_1\times M_2,\nu)
&={\textstyle
\bigoplus\limits_{p+q=r}}{1\over\nu}\left\{
\lambda\ee_p^{\prime\prime}(M_1,\lambda)
\wedge \ee_q^{\prime}(M_2,\mu)
\oplus\lambda\ee_p^{\prime\prime}(M_1,\lambda)
\wedge \ee_q^{\prime\prime}(M_2,\mu)\right.\cr
&\oplus(-1)^{p+1}
\sqrt{\lambda\mu}\ee_{p+1}^{\prime}(M_1,\lambda)
\wedge \ee^{\prime\prime}_{q-1}(M_2,\mu)\cr
&\oplus(-1)^p
\sqrt{\lambda\mu}\ee_{p-1}^{\prime\prime}(M_1,\lambda)
\wedge \ee^{\prime
}_{q+1}(M_2,\mu)\cr
&\left.\oplus\mu
\ee_p^{\prime}(M_1,\lambda)\wedge \ee^{\prime\prime}_q(M_2,\mu)
\oplus
\mu\ee_p^{\prime\prime}(M_1,\lambda)\wedge \ee^{\prime\prime}_q(M_2,\mu)
\right\}
\cr}
\no$$
This now simplifies to
$$\eqalign{\ee_r^{\prime\prime}(M_1\times M_2,\nu)
=&{\textstyle\bigoplus\limits_{p=0}^{r-1}} {\lambda\over\nu}
\ee_p^{\prime\prime}(M_1,\lambda)
\wedge \ee_{r-p}^{\prime}(M_2,\mu)
{\textstyle\bigoplus\limits_{p=0}^{r}}
\ee_p^{\prime\prime}(M_1,\lambda)
\wedge \ee_{r-p}^{\prime\prime}(M_2,\mu)\cr
&{\textstyle\bigoplus\limits_{p=1}^{r}}
(-1)^p{\sqrt{\lambda\mu}\over\nu}
\ee_p^{\prime}(M_1,\lambda)
\wedge \ee_{r-p}^{\prime\prime}(M_2,\mu)\cr
&{\textstyle\bigoplus\limits_{p=0}^{r-1}}
(-1)^{p+1}{\sqrt{\lambda\mu}\over\nu}
\ee_p^{\prime\prime}(M_1,\lambda)
\wedge \ee_{r-p}^{\prime}(M_2,\mu)\cr
&{\textstyle\bigoplus\limits_{p=1}^{r}}{\mu\over\nu}
\ee_p^{\prime}(M_1,\lambda)
\wedge \ee_{r-p}^{\prime\prime}(M_2,\mu)\cr}\no$$
\par
Some summands are now pairable off and after further minor algebraic
adjustments we find that
$$\eqalign{\ee_r^{\prime\prime}(M_1\times M_2,\nu)
=&{\textstyle\bigoplus\limits_{p=0}^{r-1}}
\left\{C_p(\lambda,\mu)\ee_p^{\prime\prime}(M_1,\lambda)
\wedge \ee_{r-p}^{\prime}(M_2,\mu)\right.\cr
&\left.\qquad\oplus
C_p(\mu,\lambda)\ee_{p+1}^{\prime}(M_1,\lambda)
\wedge \ee_{r-p-1}^{\prime\prime}(M_2,\mu)\right\}\cr
&{\textstyle\bigoplus\limits_{p=0}^{r}}
\ee_p^{\prime\prime}(M_1,\lambda)
\wedge \ee_{r-p}^{\prime\prime}(M_2,\mu)
,\qquad\hbox{where }
C_p(\lambda,\mu)={(\lambda+(-1)^{p+1}\sqrt{\lambda\mu})\over\nu}
\cr}\no$$
We can now proceed to give a much more compact formula for
$\ee_r^{\prime\prime}(M_1\times M_2,\nu)$ by first defining the
spaces $V_{r}^{\prime\prime}(\nu)$, $V_{p,r}^{\prime\prime}(\nu)$,
and $W_r^{\prime\prime}(\nu)$, $W_{p,r}^{\prime\prime}(\nu)$. We
define
\eqlabel{\vandwpp}
$$\eqalign{V_{p,r}^{\prime\prime}(\nu)=\;&
C_p(\lambda,\mu)\ee_p^{\prime\prime}(M_1,\lambda)
\wedge \ee_{r-p}^{\prime}(M_2,\mu)
\oplus\cr
&C_p(\mu,\lambda)\ee_{p+1}^{\prime}(M_1,\lambda)
\wedge \ee_{r-p-1}^{\prime\prime}(M_2,\mu)\cr
\hbox{with }\qquad V_r^{\prime\prime}(\nu)=\;&
{\textstyle\bigoplus\limits_{p=0}^{r-1}}V_{p,r}^{\prime\prime}(\nu)\cr
\hbox{and}\qquad W_{p,r}^{\prime\prime}(\nu)
=\;&\ee_p^{\prime\prime}(M_1,\lambda)
\wedge \ee_{r-p}^{\prime\prime}(M_2,\mu)\cr
\hbox{with }\qquad W_p^{\prime\prime}(\nu)=\;&
{\textstyle\bigoplus\limits_{p=0}^{r}}W_{p,r}^{\prime\prime}(\nu)\cr}\no$$
\par
Our final formula for $\ee_r^{\prime\prime}(M_1\times M_2,\nu)$ is
then
\eqlabel{\decompone}
$$\ee_r^{\prime\prime}(M_1\times M_2,\nu)=
V_r^{\prime\prime}(\nu)\oplus W_r^{\prime\prime}(\nu),\quad\nu=\lambda+\mu,\;
\lambda>0,\;\mu>0\no$$
Of course an exactly analogous formula exists for
$\ee_r^{\prime}(M_1\times M_2,\nu)$ and to give it we now make the
parallel definitions
\eqlabel{\vandwp}
$$\eqalign{V_{p,r}^{\prime}(\nu)=\;&
C_p(\lambda,\mu)\ee_p^{\prime}(M_1,\lambda)
\wedge \ee_{r-p}^{\prime\prime}(M_2,\mu)
\oplus\cr
&C_p(\mu,\lambda)\ee_{p-1}^{\prime\prime}(M_1,\lambda)
\wedge \ee_{r-p+1}^{\prime}(M_2,\mu)\cr
\hbox{with }\qquad V_r^{\prime}(\nu)=\;&
{\textstyle\bigoplus\limits_{p=1}^{r}}V_{p,r}^{\prime}(\nu)\cr
\hbox{and}\qquad W_{p,r}^{\prime}(\nu)=\;&\ee_p^{\prime}(M_1,\lambda)
\wedge \ee_{r-p}^{\prime}(M_2,\mu)\cr
\hbox{with }\qquad W_p^{\prime}(\nu)=\;&
{\textstyle\bigoplus\limits_{p=0}^{r-1}}W_{p,r}^{\prime\prime}(\nu)\cr}\no$$
and so we have
\eqlabel{\decomptwo}
$$\ee_r^{\prime}(M_1\times M_2,\nu)=
V_r^{\prime}(\nu)\oplus W_r^{\prime}(\nu),\quad\nu
=\lambda+\mu,\;\lambda>0,\;\mu>0\no$$
\par
A particularly important simplification happens to the
decompositions \docref{decompone} and \docref{decomptwo} above when,
as happens in the present paper, one can have $\mu=0$. When $\mu=0$
the spaces $\ee_r^{\prime}(M_2,0)$ and $\ee^{\prime\prime}(M_2,0)$
{\it coincide} and contain {\it harmonic} forms. We shall denote
these harmonic spaces by $\ee^H_r(M_2)$. It is then easy to check
that the decompositions above are replaced by
$$\eqalign{\ee_r^{\prime}(M_1\times M_2,\nu)&=
{\textstyle\bigoplus\limits_{p=1}^{r}}\ee_p^{\prime}(M_1,\lambda)
\wedge \ee^H_{r-p}(M_2),\quad\nu=\lambda+\mu,\;\mu=0 \cr
           \ee_r^{\prime\prime}(M_1\times M_2,\nu)&=
{\textstyle\bigoplus\limits_{p=0}^{r}}\ee_p^{\prime\prime}(M_1,\lambda)
\wedge \ee^H_{r-p}(M_2) ,\quad\nu=\lambda+\mu,\;\mu=0
\cr}
\no$$
These decompositions are essential for sections 3 and 4.
\subsec{The Product Formula}
Recalling from (\docref{taudef}) that the logarithm of the torsion
is given by the derivative at $s=0$ of $\tau(s)$ which from the
above discussion takes the form
\eqlabel{\torsions}
$$\tau(s)=-\sum_{r=0}^{m-1}\sum_{\nu}{(-1)}^r
{\Gamma_{r}''(\nu)\over{\nu}^s}\no$$
Now from our discussion above we have with $\nu=\lambda+\mu$ and
$\lambda,\mu>0$, and recalling that $\Gamma_r^{\prime\prime}=\dim
\ee_r^{\prime\prime}$ we have
$$\Gamma''_{r}(\nu)=\sum_{p=0}^{r}\Gamma''_{p}(\lambda)\Gamma''_{r-p}(\mu)
+\sum_{p=0}^{r-1}\Gamma''_{p}(\lambda)\Gamma'_{r-p}(\mu)\no$$
which by the isomorphisms identifying $\ee'_{q}\simeq\ee''_{q-1}$
gives
$$\Gamma''_{r}(\nu)=\sum_{p=0}^{r}\Gamma''_{p}(\lambda)\Gamma_{r-p}''(\mu)
+\sum_{p=0}^{r-1}\Gamma_{p}''(\lambda)\Gamma_{r-p-1}''(\mu)\no$$
Noticing that in (\docref{torsions}) if we consider a fixed
eigenvalue $\nu$ this latter decomposition implies when summed over
$r$
$$\sum_{r=0}^{d-1}{(-1)}^r\sum_{p}^{r}\Gamma''_{p}(\lambda)\Gamma_{r-p}''(\mu)
+\sum_{r=0}^{d-1}{(-1)}^r\sum_{p=0}^{r-1}
\Gamma''_{p}(\lambda)\Gamma''_{r-p-1}(\mu)\no$$
Now letting $r'=r-1$ in the second sum and setting $\Gamma''_{-1}=0$
we get on some cancellation of terms
$${(-1)}^{d-1}\sum_{p=0}^{d-1}\Gamma''_{p}(\lambda)\Gamma''_{d-p-1}(\mu)$$
Now letting $\dim M_{1}=d_1$ and $\dim M_2=d_2$, so that  $d=d_1+d_2$;
then since $\Gamma''_{p}(\lambda)=0$ for $p>(d_1-1)$, and $\Gamma''_{q}=0$
for $q>(d_2-1)$,  there is no non-zero contribution to this sum.
We conclude that the sum over degeneracies contributing to $\tau(s)$
vanishes for any non-zero $\mu$ and $\nu$.
\par
Thus the only contribution to (\docref{torsions}) comes from $\nu$
with $\mu=0$ since the other terms cancel. It is important to
observe that the terms with $\mu$ and $\nu$ non-zero cancel
eigenvalue by eigenvalue, due to the alternation in signs from
different degree forms and the isomorphisms described above. This
will be useful to our discussion of the field theory also. A little
rearrangement of this sum then shows that the sum
$\sum_{q=0}^{d_2}{(-1)}^q\Gamma_{q}(0)$ can be factored out; but
$\Gamma_{q}(0)=b_{q}$ the $q$th Betti number, this means that the
preceding sum is the Euler character of $M_2$. Thus
$$\tau(s)=\chi(M_2)\tau_{1}(s)\no$$
Taking the derivative of this at $s=0$ establishes the product
formula for the Ray-Singer torsion.
\subsec{Action Terms and Product Manifolds}
We employ an important and  useful notation in the action to denote
the degree of the forms appearing therein:
$$\eqalign{&f_r\qquad \hbox{denotes a form on $M$
of degree } r\cr &\bar f_r \qquad
\hbox{denotes a form on $M$ of degree } (n-r)\cr} \qquad
\hbox{where } n=\dim M\no$$
This allows us to develop a much clearer notation for the BRST field
theory below; in particular the dimension $n$ of $M$ does not
explicitly appear and expressions such as $\bar g_r\wedge f_r$ are
immediately recognisable as volume forms.
\par
To actually compute the functional integral for the torsion  we have
only three types of contribution  to the action  to consider, namely
\eqlabel{\threetypes}
$$\int_M  f_{r-1}^{\prime\prime}\wedge d\bar f_r^{\prime\prime},
\qquad \int_M \bar f_{r-1}^\prime\wedge d^* f_r^\prime, \qquad
\hbox{and }\int_M\bar f_r^\prime\wedge d^*d f_r^{\prime\prime}\no$$
However these are expressible using natural inner products induced
by the metric on $M$.
\par
In general for two $r$-forms $\omega$ and $\nu$ on $M$ we  define
their  inner product by
$$<\omega, \nu>=\int_M tr(\omega\wedge *\nu)\no$$
In what follows we shall relieve the notation by omitting the trace
on the RHS and the wedge between the forms.
\par
Turning now to the first of our three contributions above we
have\footnote*{For keeping abreast of the  signs in some of these
formulae it is useful to note that the operators
$*:\Omega_r(M)\rightarrow\Omega_{n-r}(M)$ and
$d^*:\Omega_r(M)\rightarrow\Omega_{r-1}(M)$ satisfy
$*^2=(-1)^{r(n-r)}$ and $d^*=(-1)^{nr+n+1}*d*$.}
$$\eqalign{\int_M f_{r-1}^{\prime\prime} d\bar f_r^{\prime\prime}&=
{(-1)}^{(n+1)(r-1)}\int_M f_{r-1}^{\prime\prime} *^2 d\bar
f_r^{\prime\prime}\cr &={(-1)}^{(n+1)(r-1)}<f_{r-1}^{\prime\prime},
*d\bar f_r^{\prime\prime}>\cr}\no$$
But we know from \docref{isomor} that $*d$ gives rise to an
isomorphism between the spaces $\ee_r^{\prime\prime}(M,\nu)$ and
$\ee_{r+1}^{\prime}(M,\nu)$ for each $\nu$. To make use of that here
we use the decomposition
$$\ee_r^{\prime\prime}(M)=
{\textstyle\bigoplus\limits_{\nu}}\,\ee_r^{\prime\prime}(M,\nu)\no$$
Now we let $\{e^{\prime\prime}_r(M,\nu,i)\}$ denote an orthonormal
basis for $\ee_r^{\prime\prime}(M,\nu)$ ($i$ labels the degeneracy)
and  write, remembering that $\bar f$ denotes an $(n-r)$-form,
\eqlabel{\decomp}
$$\eqalign{\bar f_r^{\prime\prime}& =\sum_{\nu,i}
\bar c_r^{\prime\prime}(\nu,i)e^{\prime\prime}_{n-r}(M,\nu,i)\cr
f_{r-1}^{\prime\prime}&=\sum_{\nu,j} c_{r-1}^{\prime\prime}(\nu,j)
e_{r-1}^{\prime\prime}(M,\nu,j)\cr}\no$$
where $\bar c_r^{\prime\prime}(\nu,i)$ and
$c_{r-1}^{\prime\prime}(\nu,j)$  are constants. But using the
isomorphisms \docref{isomor} and \docref{poincare} we calculate that
$$*d\bar f_r^{\prime\prime}=\sum_{\nu,i} \bar
c_r^{\prime\prime}(\nu,i)\sqrt{\nu} e^{\prime\prime}_{r-1}(M,\nu,i)\no$$
which means that
\eqlabel{\termone}
$$\eqalign{\int_M f_{r-1}^{\prime\prime} d\bar f_r^{\prime\prime}
&= {(-1)}^{(n+1)(r-1)}\kern-7pt\sum_{\nu_1,i,\nu_2,j}
c_{r-1}^{\prime\prime}(\nu_{1},j) \bar
c_r^{\prime\prime}(\nu_{2},i)\sqrt{\nu}
<e^{\prime\prime}_{r-1}(M,\nu_{1},j),e^{\prime\prime}_{r-1}(M,\nu_{2},i)>\cr
&={(-1)}^{(n+1)(r-1)}\sum_{\nu,i} c_{r-1}^{\prime\prime}(\nu,i)\bar
c_r^{\prime\prime}(\nu,i)\sqrt{\nu} \cr}\no$$
Similarly, and in an analogous notation, we calculate that
\eqlabel{\termstwothree}
$$\eqalign{\int_M \bar f_{r-1}^\prime d^*
f_r^\prime&={(-1)}^{(n+1)(r-1)} \sum_{\nu,i}\bar
c_{r-1}^{\prime}(\nu,i)c_r^{\prime}(\nu,i)\sqrt{\nu}\cr \hbox{and
}\int_M\bar f_{r}^\prime d^*d f_r^{\prime\prime}&=
{(-1)}^{(n+1)(r-1)}\sum_{\nu,i}\bar c_{r}^{\prime}(\nu,i)
c_r^{\prime\prime}(\nu,i)\nu\cr}\no$$
Hence \docref{termone} and \docref{termstwothree} give our
evaluations of the three generic action terms of \docref{threetypes}
above.
\par
Let us now consider what happens on a product manifold,
$M=M_{1}\times M_{2}$. In this case the spaces
$\ee_r^{\prime}(M,\nu)$  and $\ee_r^{\prime\prime}(M,\nu)$ decompose
further as described above. It is convenient for our purposes to
separate off the harmonic contribution to $M_{2}$ from the
remainder. Thus, denoting the orthonormal basis for the space
$\ee^H_j(M_2)$ of harmonic $j$-forms by $\{\e_{j}^a\}$, we have
$$f_{r}=\widetilde f_{r}+\sum_{a=1}^{b_{j}}
\sum_{j=0}^{min(r,d_2)}f_{r-j}^a\e_{j}^{a},\qquad d_2=\dim M_2$$
The term $\widetilde f_{r}$ which denotes the field remaining after
separating off the harmonic contribution, further decompose into
$f_r^{\prime}+f_r^{\prime\prime}$ of the form (\docref{decomp})
where $\mu>0$. But, we  have shown that a basis for such forms on a
product manifold is provided by \docref{vandwpp} and
\docref{vandwp}, thus $\widetilde f_r^{\prime\prime}$ further
decomposes as
\eqlabel{\fvw}
$$\tilde f_r^{\prime\prime}=
\sum_{p=0}^{r-1}f^{V_{r,p}^{\prime\prime}}_{r}
+\sum_{p=0}^{r}f^{W_{r,p}^{\prime\prime}}_{r}$$
Similarly $\f_{r}$ decomposes as
$$\f_{r}=\tfb_{r} + \sum_{a=1}^{\bar b_j}
\sum_{j=0}^{min(r,d_2)}\f_{r-j}^{a}\bar\e_{j}^{a}, \qquad
d_2=\dim M_2$$

where this time the sum over $a$ is a sum over an orthonormal basis
for the space $\ee^H_{d_2-j}(M_2)$ whose dimension is $(d_2-j)=\bar
b_j$; $\bar b_j$ being the $(d_2-j)$-th Betti number. For an
orientable manifold $\bar b_j= b_j$ and we will assume this to be
the case though little modification is necessary to treat the more
general case.  Were both factors to have harmonic contributions the
above formulae generalize in a symmetrical fashion, this case is not
of interest here. \par Now, retaining only terms which do not
integrate to zero, our three fiducial terms have the decomposition
\eqlabel{\dcmpfbdf}
$$\f_{r-1}d^*f_{r}=\tfb_{r-1}d^*\widetilde f_{r}
+\sum_{j=0}^{r}\sum_{a=1}^{b_{j}}\f_{r-j-1}^{a}d^*f_{r-j}^{a}\e^{a}\no$$
where $\e^{a}=\bar\e_j^a\e_j^a$ and so is a volume form on $M_2$.
\par
Similarly
\eqlabel{\dcmpfdfb}
$$f_{r-1}d\f_{r}=\widetilde f_{r-1}d\tfb_{r}
+\sum_{j=0}^{r}\sum_{a=1}^{b_j}f_{r-j-1}^{a}d\f_{r-j}^a\e^{a}\no$$
and
\eqlabel{\dcmpdiag}
$$\f_{r}d^*df_{r}=\tfb_{r}d^*d\widetilde f_{r}
+\sum_{j=0}^{r}\sum_{a=1}^{b_j}\f_{r-j}^{a}d^*df_{r-j}^{a}\e^{a}\no$$
We will have occasion to  use these formulae in the next sections.
Now we turn to  the matter of the BRST quantization which we will
need for the computation of the torsion. \beginsection{BRST
Quantisation: The Gauge Fermion Construction} We wish to calculate
the partition function associated with the functional integral over
the action $S[f_{n}]$ \eqlabel{\partnfn}
$$Z=\int\mu[f_{n}]e^{-S[f_{n}]}\no$$ where the  action $S[f_n]$ is
invariant under the local gauge transformation
\eqlabel{\trans}
$$f_{n} \r f_{n} + d \omega_{n-1}\no$$
The functional measure $\mu[f_{n}]$ must contain gauge fixing delta
functionals to ensure that the integration is performed only over
one of the set of $f_{n}$ which are equivalent under such a gauge
transformation. An action that will be of special concern to us in
the context of topological field theory and the Ray-Singer torsion
is \eqlabel{\omn} $$S[f_{n}]=i\int_{M}f_{n}d f_{n}\no$$ where
$f_{n}$ is a matrix valued $n$-form. However since the methods of
this section are not particular to the action $S[f_n]$ chosen we
will not restrict ourselves to any particular action for the
present. Our focus is on the construction of the the measure
$\mu[f_n]$.
\par
This measure  is most simply constructed by extending the set of
fields being integrated over, in such a way that on integrating out
the additional fields the result yields the desired measure
$\mu[f_{n}]$. The necessary extended set of fields and resulting
action are obtained by replacing the original transformation
(\docref{trans}) by a BRST transformation $$\delta f_{n}=d
f_{n-1}\no$$ The transformation $\delta$ is of Fermionic character
in that the field $f_{n-1}$ has the opposite statistics to those of
the field $f_{n}$. Notice that the original action is invariant
under this BRST transformation despite its Fermionic nature, so that
$$\delta S[f_{n}]=0\no$$
A set of transformation laws must similarly be given for the
additional fields. The `ghost' field $f_{n-1}$ is the first member
of this extended set of additional fields. The method we will follow
for the construction of the auxiliary fields and action is that of
Batalin and Vilkovisky \ref{7,8}. In this method the gauge fixing is
performed at the level of a ``gauge Fermion'' $\Psi$ whose BRST
variation $\delta\Psi$ gives the necessary addition to the original
action \docref{omn}. The BRST action is then
\eqlabel{\omnpsi}
$$\hat S=S[f_{n}]+\int_{M}\delta\Psi\no$$
If we ensure that $\delta^2=0$ then the original gauge invariance of
the action (\docref{omn}) has been replaced by a BRST invariance of
the action (\docref{omnpsi}). \par We now proceed with the
construction of the gauge Fermion $\Psi$ for the gauge fixing of the
field $f_n$. It is convenient to construct the ghost Fermion as a
triangle and to start from its bottom which we declare to be of
level zero. For this purpose it is convenient to introduce some
important notation: We will label our fields by $f_{(i,j)}$ and
$\f_{(i,j)}$ where
$$\eqalign{&f_{(i,j)}\qquad\qquad\hbox{denotes a form of degree $i$
and level $j$}\cr \hbox{and }\quad &\bar f_{(i,j)}\qquad\qquad
\hbox{denotes a form of degree $(n-i)$ and level $j$.}}\no$$
Summarising: the  first suffix indicates the degree of the form and
the second indicates the ghost level (as distinct from ghost
number\footnote*{Actually the ghost number of $f_{(i,j)}$ is $(n-j)$
and that of the anti-ghost $\f_{(i,j)}$ is $-(n-j)$.
This can be deduced from the requirement that the action should have
ghost number zero and the fact that $\delta$ increases ghost number by
one.}).
Our notation will ensure that details of the underlying manifold
play a minimal r\^ole. This is convenient since the construction is
local and algebraic. \subsec{The Gauge Fermion and its Variation}
The pattern that emerges for the gauge Fermion  may be compactly
written down for the general term, at  level $k$ for the  gauge
Fermion. It is \eqlabel{\levelkgf}
$$\Psi_{k}=\sum_{j=0}^{[k/2]}\Psi_{(k-2j,k)},
\qquad\hbox{with }[k/2]=\hbox{the integer part of $k$}\no$$
where
\eqlabel{\basicterm}
$$\Psi_{(k-2j,k)}=\bar
f_{(k-2j-1,k-1)}d^*f_{(k-2j,k)}+f_{(k-2j-2,k)}d\bar f_{(k-2j-1,k-1)}\no$$
Note that the expression \docref{basicterm} for $\Psi_{(k-2j,k)}$
has two terms: the first term contains  a gauge fixing for the field
$f_{(k-2j,k)}$ but in the process introduces a new field
$\f_{(k-2j-1,k-1)}$, this new field also requires gauge fixing
which, one notes,  is done by the second term; however this second
term contains a further new field $f_{(k-2j-2,k)}$ which needs
dealing with. But this latest field is of the same form of the
original one except that its first suffix is smaller by two. Hence
we can repeat the process just described until the first suffix
reaches zero.
\par
We use the subscripts $(k-2j,k)$ for $\Psi_{(k-2j,k)}$ since they
are also those of the leading field being gauge fixed by this
structure, and all information is recoverable from these subscripts.
\par
The pattern of the gauge Fermion can be summarized in the
following triangular diagram
\par\vskip 1.5\baselineskip
$$\matrix{
\ddots&\qquad &\ddots&\qquad &\ddots\hfill&\qquad  &\ddots\hfill \cr
& & & & & & \cr & & & & & & \cr & & & & & & \cr
& & \f_{(2,2)}d^*f_{(3,3)}& &f_{(1,3)}d\f_{(2,2)}&
&\f_{(0,2)}d^*f_{(1,3)}\cr
& & & & & & \cr
& & & & & & \cr
& & & & & & \cr
& & & & \f_{(1,1)}d^*f_{(2,2)}& &f_{(0,2)}d\f_{(1,1)}\cr
& & & & & & \cr
& & & & & & \cr
& & & & & & \cr
& & & & & & \f_{(0,0)}d^*f_{(1,1)}\cr }$$
\par\vskip 2\baselineskip\penalty 10000
\centerline{{\bf Fig 1:} {\it The Gauge Fermion Triangle}}
\par\vskip 2\baselineskip We now come to the BRST variation of this
whole structure. The BRST variations of the corresponding fields are
$$\eqalign{\delta f_{(k,k)}&=df_{(k-1,k-1)},\cr \delta
f_{(k-2j,k)}&=f_{(k-2j,k-1)}, \qquad \delta f_{(k-2j,k-1)}=0,\qquad
j\ne0\cr \delta\f_{(k-2j,k-1)}&=\f_{(k-2j-1,k)}, \qquad
\delta\f_{(k-2j-1,k)}=0,\qquad \forall j \cr}\no$$ We note that
$\delta$ commutes with $d$ and $d^*$ and $\delta^2=0$.

The variation of $\Psi_{k}$ itself comprises two sets of terms: one
set of ghost level $k$ the other of ghost level $k-1$. This
splitting into two levels arises  from the Leibnitz rule applied to
the product of a level $k$ field with a level $k-1$ anti-field; one
can see how it operates by following the downward and upward arrows
of the variation triangle of Fig. 2 below.
\par
It is convenient to organize the resulting fields into groups of a
given level on this basis. The resulting variation is
\eqlabel{\levelkvar}
$$\delta\Psi_{k}=\sum_{j=0}^{[k/2]}{\cal F}_{(k-2j,k)}
+\bar f_{(k-1,k-1)}d^*df_{(k-1,k-1)}
+\sum_{j=0}^{[k/2]}\bar{\cal F}_{(k-2j-1,k-1)}
\no$$
where the basic structures that appear are
$${\cal F}_{(k-2j,k)}=
\f_{(k-2j-1,k)}d^*f_{(k-2j,k)}+{(-1)}^{F}f_{(k-2j-2,k)}d\f_{(k-2j-1,k)}\no$$
and
$$\bar{\cal F}_{(k-2j,k)}
=f_{(k-2j-1,k)}d\f_{(k-2j,k)} +{(-1)}^{F}\f_{(k-2j-2,k)}d^*f_{(k-2j-1,k)}\no$$
where $F$ is the Fermion number of the field appearing immediately
to its right. These are of similar form to $\Psi_{(k-2j,k)}$,
however ${\cal F}_{(k-2j,k)}$ and  $\bar{\cal F}_{(k-2j,k)}$ both
consist entirely of fields at the same level $k$.
\par
Note that in the expression \docref{levelkvar} for $\delta\Psi_{k}$
we have separated out, as special, the  diagonal term,
$f_{(k-1,k-1)}$. This field will acquire its gauge fixing from the
level $k-1$ ghost Fermion, $\Psi_{k-1}$, just as the field
$f_{(k,k)}$ is gauge fixed by the first sum  of
(\docref{levelkvar}).
\par
The structure which emerges as a result of the BRST transformation
is a new triangle which can be described as follows: one starts with
the original triangle of Fig. 1. and, after BRST variation, two
triangles appear: one displaced upwards to the right the other
displaced downwards to the right. These two triangles partially
overlap to form a new triangle containing a distinguished diagonal.
This structure, where we include $\Psi$ and $\delta\Psi$ in the same
triangle, is illustrated in Fig. 2 below.
\par\vskip 1.5\baselineskip\noindent
$\matrix{
& & & &\ddots&  &  &\ddots& &\ddots \cr
& & & &&  &  && & \cr
& & & & &\nearrow & & &\nearrow &\cr
& & & &\f_{(1,1)}d^*f_{(2,2)}& & &f_{(0,2)}d\f_{(1,1)}& &\cr
& & & & &\searrow & & &\searrow &\cr
& & & & & &\f_{(1,1)}d^*df_{(1,1)} & &
&f_{(0,1)}d\f_{(1,1)}+\f_{(0,1)}d^*f_{(1,1)}\cr
& & & & & & & &\nearrow &\cr
& & & & & & & \f_{(0,0)}d^*f_{(1,1)}&\cr
& & & & & & & &\searrow &\cr
& & & & & & & & &\f_{(0,0)}d^*df_{(0,0)}\cr
}$
\par\penalty 10000\vskip 1\baselineskip plus 0.6\baselineskip minus
0.6\baselineskip
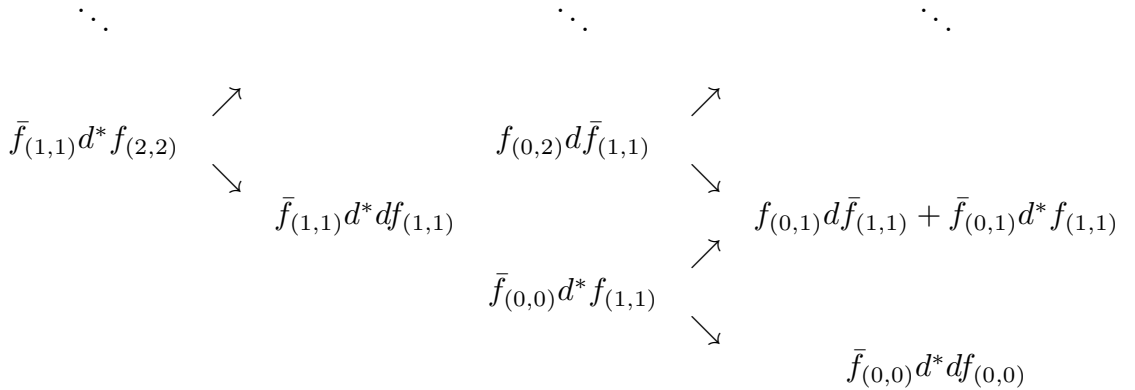
\centerline{\bf {Fig 2:} {\it The gauge Fermion triangle and its variation}}
\par\vskip \baselineskip
\par\noindent
The south east arrow represents the variation of an un-barred field,
the north east arrow that of the barred field. The variation of a
barred field {\it increases} its second index by one while the
variation  of an unbarred  field {\it decreases} its second index by
one.  We see the variation of diagonal elements is special as the
triangle is upper diagonal. Similarly if the gauge Fermion triangle
terminates the top terminal row is special having no associated
downward partner.
\par
We are now in a position to complete the gauge fixing for the field
$f_{(n,n)}$ in the action  $S[f_{(n,n)}]$. The gauge Fermion
required is simply
$$\Psi = \sum_{k=1}^{n}\Psi_{k}\no$$
and is a sum of all rows of the triangle up to level $n$. The
resulting BRST invariant gauge fixed Lagrangian in terms of our
compact notation is just
$$\hat {\cal L}= {\cal L}_{n}+{\cal L}$$
where
$${\cal L}=\sum_{j=0}^{[n/2]}{\cal F}_{(n-2j,n)}
+\sum_{k=0}^{n-1}{\cal L}_{k}\no$$
and
$${\cal L}_{k}=\f_{(k,k)}d^*df_{(k,k)}
+\sum_{j=0}^{[k/2]}\left(\bar{\cal F}_{(k-2j,k)}
+{\cal F}_{(k-2j,k)}\right)\no$$
and ${\cal L}_{n}$ is the the original Lagrangian \docref{omn} for
the field $f_{(n,n)}$.
\par
The boundary contribution, i.e. level $n$ contribution to ${\cal L}$
takes the form
\eqlabel{\boundcont}
$$\sum_{j=0}^{[n/2]}{\cal F}_{(n-2j,n)}\no$$
and does not have corresponding $\bar{\cal F}$ contributions.
\par
Our next task is to integrate out all the fields $f_{(i,j)}$ and
$\f_{(i,j)}$ to obtain the measure $\mu[f_{(n,n)}]$. The first
observation is that none the Lagrangians ${\cal L}_{k}$ `interact'
with each other and we can therefore integrate them out separately.
Next we observe that if we use the decomposition described earlier
into primed and double primed forms ($d\phi'=0$ and $d^*\phi''=0$)
then we find each term in the Lagrangian involves a distinct field,
i.e. we can write
$$\f_{(k,k)}d^*df_{(k,k)}=\f_{(k,k)}''dd^*f_{(k,k)}'\no$$
$${\cal F}_{(k-2j,k)}=\f_{(k-2j-1,k)}' d^*f_{(k-2j,k)}'
+{(-1)}^{F}f_{(k-2j-2,k)}'' d\f_{(k-2j-1,k)}''\no$$
and
$$\bar{\cal F}_{(k-2j,k)}=f_{(k-2j-1,k)}'' d\f_{(k-2j,k)}''
+{(-1)}^{F}\f_{(k-2j-2,k)}' d*f_{(k-2j-1,k)}' \no$$
\par
Up to this point we have been careful not to specify the statistics
of the fields, other than to remark that the statistics alternate
with the level; in addition all the terms of  a given level have the
same statistics. It is then clear that the statistics of all fields
are determined by the choice made for the apex field $f_{(0,0)}$. We
make such a choice and assign the apex statistics first (though
conventionally one would start by specifying the statistics of the
field $f_{(n,n)}$)
\par
We will  consider in turn both options for the statistics of these
fields, but begin with the Fermionic case; this being the one  which
occurs when these fields arise in the gauge fixing of a $1$-form
field such as in QED.
\par
Integrating out just these fields at the apex then gives
$\Det(d^*d_{0})$ which we simply denote by $X_{0}$, the subscript
denoting the fact that this object is associated with an operator
acting on zero forms.
\par
Let us now consider the level $k$ Lagrangian, ${\cal L}_{k}$. The
leading term in this Lagrangian is
$$\f_{(k,k)}''d^*df_{(k,k)}'\no$$
and it will involve Fermionic fields if $k$ is even and Bosonic
fields if $k$ is odd. Integrating out this term then gives
\eqlabel{\Xkdef}
$$X_{k}^{p_k},\qquad \hbox{where } X_{k}=\Det(d^*d_{k}),\quad \hbox{and }
p_k={(-1)}^k\no$$
Now if we integrate out the  term
$${\cal F}_{(k-2j,k)}=
\f_{(k-2j-1,k)}'d^*f_{(k-2j,k)}'
+{(-1)}^{F}f_{(k-2j-2,k)}''d\f_{(k-2j-1,k)}''\no$$
we obtain
$$\Det(d^*_{k-2j})^{p_k}\Det(d_{k-2j-2})^{p_k}\no$$
By the isomorphisms discussed previously,  we obtain
$$\Det(d^*_{k-2j})=X_{k-2j-1}^{1/2}\qquad
\Det(d_{k-2k-2})=X_{k-2j-2}^{1/2}\no$$
Therefore the contribution from integrating out ${\cal F}_{(k-2j,k)}$ is
$$X_{k-2j-1}^{p_k/2}X_{k-2j-2}^{p_k/2}\no$$
Similarly integrating out the fields in $\bar{\cal F}_{(k-2j-1,k)}$ yields
$$\Det(d^*_{k-2j-1})^{p_k}\Det(d_{k-2j-1})^{p_k}
 =X_{k-2j-2}^{p_k/2}X_{k-2j-1}^{p_k/2}\no$$
We see that both ${\cal F}$ and $\bar{\cal F}$ terms give similar
contributions when integrated out. In effect as a consequence of the
field theory Poincar\'e duality is automatically implemented.
\par
The  result therefore of integrating out all fields contributing to
${\cal L}_{k}$ is
$$ \matrix{&X_{0}X_{1}X_{2}\dots X_{k} &\qquad \hbox{for $k$ even}\cr
& &\cr
&(X_{0} X_{1} X_{2} \dots X_{k})^{-1} &\qquad \hbox{for $k$ odd}\cr}\no$$
with the statistics of $f_{(0,0)}$ Fermionic.
\par
Now integrating out all fields in the Lagrangians
$\sum_{i=0}^{k}{\cal L}_{i}$ yields
$$ \matrix{&X_{0}X_{2}X_{4}\dots X_{k} &\qquad \hbox{for $k$ even}\cr
\cr
&(X_{1} X_{3}\dots X_{k})^{-1} &\qquad \hbox{for $k$ odd}\cr}\no$$
\par
There  remains only to consider the final boundary term
\docref{boundcont}. These fields when integrated out give the
contribution
$$\matrix{&X_{0}^{1/2}X_{1}^{1/2}X_{2}^{1/2}\dots X_{n-1}^{1/2}
&\qquad \hbox{for $n$ even }\cr
\cr
&(X_{0}^{1/2}X_{1}^{1/2}X_{2}^{1/2}\dots X_{n-1}^{1/2})^{-1}
&\qquad \hbox{for $n$ odd }\cr}\no$$
\par
Integrating out all of the fields associated with the gauge fixing
of $f_{(n,n)}$ yields the complete measure
\eqlabel{\measure}
$$\mu[f_{(n,n)}]=\mu_{(n,n)}\delta[d^*f_{(n,n)}]\no$$
with
\eqlabel{\brstdet}
$$\mu_{(n,n)}={X_{0}^{1/2}X_{2}^{1/2}\dots X_{2j}^{1/2}\dots
\over X_{1}^{1/2}X_{3}^{1/2}\dots X_{2j+1}^{1/2}\dots }
\hbox{ \qquad for all $n$}\no$$
where the dots terminate with the last term $X_{n-1}$ and the field
$f_{(n,n)}$ is Fermionic for $n$ even and Bosonic for $n$ odd, since
we have chosen $f_{(0,0)}$ to be Fermionic.---If the statistics of
the field $f_{(n,n)}$  are opposite to those above then the
corresponding result can be read off by replacing $X_{k}$ by
$X_{k}^{-1}$ in all expressions above---We emphasize that the above
considerations leading to the result (\docref{brstdet}) are quite
general and do not rely on the form of the action to be gauge fixed;
they rely only on the fact that the field to be gauge fixed is an
$n$-form, and that the action is invariant under the addition of a
closed form.   The gauge Fermion itself is formally infinite and can
be considered to be a Universal Gauge Fermion $\Psi$, the finite
dimensionality of the manifold choosen and the field to be gauge
fixed determine the relevant portion of the gauge Fermion for a
particular problem.
\subsec{The Gauge Fermion on Product Manifolds}
Our next task is to analyse this structure on a product manifold
$M_1\times M_2$ where $M_2$  admits harmonic forms. As mentioned
above there are three separate types of term to analyse. If we focus
on the gauge Fermion only two of these enter. However, we have to be
a little careful to ensure that our notation does not become
degenerate. Such a  problem arises now since there are new forms at
level $k$ which, if we track them using the labels of the product,
will acquire the same labels. What  saves us is that fields
acquiring the same subscripts will have different harmonic labels.
No confusion should then arise once this is kept in mind.
\par
Let us now consider  the level $k$ gauge Fermion given by
(\docref{levelkgf}) on a product manifold where we separate off the
harmonic contributions from $M_2$. The decomposition
(\docref{dcmpfbdf}) and (\docref{dcmpfdfb}) are the relevant ones
and give
$$\Psi_{(r,k)}=\widetilde\Psi_{(r,k)}
+\sum_{j=0}^{r}\sum_{a=1}^{b_j}\Psi_{(r-j,k)}^a\bar\e_{j}^{a}\e_{j}^{a}$$
The term $\widetilde\Psi_k$ further decomposes using \docref{fvw}
and the orthogonality of such fields. We will not, however, pursue
this here. Now noting that when the first index becomes zero
$\Psi^a_{(r-j,k)}$ vanishes we can regroup the level $k$ gauge
Fermion into a triangle graded by Betti number, thus
$$\Psi_{k}=\widetilde\Psi_{k}+\sum_{j=0}^{k}\sum_{a=1}^{b_j}
\Psi_{k-j}^{a}(k)\bar\e_{j}^{a}\e_{j}^{a}$$
where we include the bracketed $k$ to indicate that these terms are
arising from the level $k$ gauge Fermion, and to avoid a degeneracy
of notation.
\par
We further simplify our notation by using a repeated index
convention for the sums over harmonic forms thus
$$\Psi=\widetilde\Psi+\Psi^{a}\e^{a}$$
The sum is understood to be over all harmonic forms and
$\e^{a}=\bar\e^a_j\e^a_j$ for $j$-forms.  The gauge Fermions
$\Psi^{a}$ are understood to terminate when the fields comprising
them do not exist. Thus in the gauge fixing of an $n$-form on $M$
the gauge Fermion associated with $j$-forms on $M_2$ will terminate
at the level $k$ for which $(n-j-k)$ reaches 1.
\par
The variation of the gauge Fermion can be similarly organized as
before to give
$${\cal L}=\widetilde{\cal L}+{\cal L}^{a}\e^a$$
where ${\cal L}^a$ are the results of varying the gauge Fermions
$\Psi^a$, the superscript $a$ runs over all allowable Betti numbers;
these will depend on the field being gauged fixed, and the  manifold
dimensions $d_1$ and $d_2$.
\par
The combined action remaining once the harmonic contribution on $M_{2}$
has been extracted is then
$$\widetilde{\cal L}=\sum_{j=0}^{[n/2]}\widetilde{\cal F}_{(n-2j,n)}
+\sum_{k=0}^{n-1}\widetilde{\cal L}_{k}\no$$
It  can then be integrated as before to give
\eqlabel{\tldbrstdet}
$$\widetilde\mu_{(n,n)}
={\widetilde X_{0}^{1/2}\widetilde X_{2}^{1/2}\dots
\widetilde X_{2j}^{1/2}\dots
\over \widetilde X_{1}^{1/2}\widetilde X_{3}^{1/2}\dots
\widetilde X_{2j+1}^{1/2}\dots }
\hbox{ \qquad for all $n$}\no$$
where $\widetilde X_{i}=\widetilde\Det(d^*d_{i})$ denotes the
determinant  of $d^*d_{i}$ on the orthogonal complement to the
harmonic forms on $M_2$.
\par
Finally examining  the harmonic sector we find on integrating out
the fields in ${\cal L}^e\e^a$ the contribution
\eqlabel{\harmonic}
$$\eqalign{\mu^H_{(n,n)}&
=\prod_{k=0}^{n-1}X_{k}(M_1)^{p_k\sigma_k}\qquad\qquad
\hbox{where }\qquad \sigma_k={1\over 2}\sum_{j=0}^{n-k-1}{(-)}^{j}b_j
\cr
\hbox{and } &X_k(M_1)\qquad\hbox{ is (\docref{Xkdef})
restricted $M_1$.}\cr}\no$$
\beginsection{Topological Field Theories}
Let us now consider a topological field theory, by choosing the
starting Lagrangian to be $f_{(n,n)}df_{(n,n)}$. Note that Stokes'
theorem shows that this Lagrangian will integrate to {\it zero} if
the field $f_{(n,n)}$ is {\it Bosonic} for $n$ {\it even} or {\it
Fermionic} for $n$ {\it odd}. Hence we are naturally led to consider
$f_{(n,n)}$ to have statistics consistent with the descendant BRST
field $f_{(0,0)}$ being Fermionic.
\par
The field term $\f_{(n,n)}df_{(n,n)}=f_{(n,n)}'df_{(n,n)}'$
integrates to give
$$\matrix{&X_{n}^{1/4} &\qquad  \hbox{for  $n$ even}\cr
&&\cr
&X_{n}^{-1/4} &\qquad  \hbox{for $n$ odd} \cr}\no$$
When included with  the contribution from BRST gauge fixing the
partition function is
$$\eqalign{
Z_{n}&={X_{0}^{1/2}X_{2}^{1/2}\dots
\over X_{1}^{1/2}X_{3}^{1/2}\dots }X_{n}^{1/4}\qquad
\hbox{ for $n$ even}\cr
&={X_{0}^{1/2}X_{2}^{1/2}\dots
\over X_{1}^{1/2}X_{3}^{1/2}\dots }X_{n}^{-1/4}\qquad
\hbox{for $n$ odd } \cr
&=\prod_{k=0}^{n-1}X_{k}^{p_{k}/2}X_{n}^{p_{n}/4}
\qquad \hbox{where $p_{k}={(-1)}^{k}$}
\cr}\no$$
The logarithm of this can be expressed in terms of zeta functions by
recalling that
$$\ln X_{k}=-\zeta_{d^*d_{k}}'(0)\no$$
thus
$$\eqalign{\ln Z_{n}&
=-{1\over2}\sum_{k=0}^{n-1}{(-1)}^{k}\zeta_{d^*d}'(0)
+{1\over4}{(-1)}^{n+1}\zeta_{d^*d_{n}}'(0)\cr
&={1\over4}\ln T \hbox{\qquad from our previous discussion. }}\no$$
We note that the partition function of our alternating sequence of
Fermionic and Bosonic actions always yields the torsion to the same
power $1/4$.
\par
This latter identification of the partition function with the
quarter power of the torsion relies on regulating the expressions
$X_{k}$. We have adopted the zeta function procedure of the
proceeding discussion. We emphasize however that we are not
restricted to this regularisation procedure. In fact,  by noting the
close relationship of the zeta function and the heat kernel, it is
possible to use any covariant regularisation at the lever of the
heat kernel. A natural one to consider from the point of view of
physics is a lattice regulator. This could be implemented by working
on a simplicial lattice. The preceding analysis will go through
without modification, provided the equivalent objects are given
their natural lattice interpretation. This is then equivalent to
M\"uller's proof of the equivalence of the combinatorial and lattice
torsions.
\par
Let us mention some further points regarding the partition function
in general before proceeding to a derivation of the product formula.
The first observation is that if we wish to consider a topological
action in {\it even} dimensions we need to begin with an alternative
to $$\int_Mf_{(n,n)}df_{(n,n)}\no$$ since this action only exists
for odd dimensional manifolds.
\par
The simplest possibility is to consider an action made from two
distinct fields say $f_{(k,k)}$ and $g_{(l,l)}$. The topological
action we consider is $g_{(l,l)}d f_{(k,k)}$ and ${\rm dim} M =
(k+l+1)$.  Of course this action can be considered in arbitrary
dimensions and for both statistics for the respective fields
provided they have the same statistics. From our work above these
fields have separate  gauge Fermions for their gauge fixing, which
do not mix. The statistics of $f_{(0,0)}$ and $g_{(0,0)}$ are not
independent but are determined by  the dimension of the manifold
since to be able to form an action from the two fields $f_{(k,k)}$
and $g_{(l,l)}$ dictates that both of these fields have the same
statistics; but since the two gauge Fermion triangles meet, there is
a constraint.
\par
For simplicity let us consider the case of an even dimensional
manifold of dimension $2n$ and pick the starting fields to be
$f_{(n,n)}$ and $g_{(n-1,n-1)}$. Since the total height of the $f$
triangle is one more than that of the $g$ triangle it is clear that
the fields $f_{(0,0)}$ and $g_{(0,0)}$ must have opposite
statistics. If we take the case when $f_{(0,0)}$ has Fermionic
statistics we obtain the contribution from the $f$ triangle as in
(\docref{brstdet}) and that from the $g$ triangle is the inverse of
this but terminating one term sooner. For definiteness   let us take
$n=2m +1$ in which case $f_{(n,n)}$ has the opposite statistics to
$f_{(0,0)}$, and is a now a Boson. Thus we have  a contribution to
the partition function of
$${X_{0}^{1/2}X_{2}^{1/2}\dots X_{2m}^{1/2}
\over X_{1}^{1/2}X_{3}^{1/2}\dots X_{2m-1}^{1/2}}
\hbox{ \qquad for the $f$ triangle}\no$$
and
$${X_{1}^{1/2}X_{3}^{1/2}\dots X_{2m-1}^{1/2}
\over X_{0}^{1/2}X_{2}^{1/2}\dots X_{2m-2}^{1/2}}
\hbox{ \qquad for the $g$ triangle}\no$$
Integrating out the original action gives $X_{2m+1}^{-1/2}$ which by
shifted Poincar\'e duality is equal to $X_{2m}^{-1/2}$.  Taking the
product of all this gives the partition function $Z_n$ so that we
have
$$Z_n={X_{0}^{1/2}X_{2}^{1/2}\dots X_{2m}^{1/2}
\over X_{1}^{1/2}X_{3}^{1/2}\dots X_{2m-1}^{1/2}}
{X_{1}^{1/2}X_{3}^{1/2}\dots X_{2m-1}^{1/2}
\over X_{0}^{1/2}X_{2}^{1/2}\dots X_{2m-2}^{1/2}}
X_{2m}^{-1/2}=1\no$$
We see that the partition function is {\it automatically} unity.
This agrees, as it must, with the torsion which is  also unity in
even dimensions.
\par
In the case of an odd dimensional manifold of dimension $2n+1$ with
starting action chosen to be
$$\int_M g_{(n,n)}df_{(n,n)}\no$$ the fields $f_{(0,0)}$ and $g_{(0,0)}$
will now have the same statistics and the contribution from each
BRST triangle will be identical. The contribution from the original
action will then be
$$X_{n}^{\pm 1/2}\no$$
plus for Fermionic fields and minus for Bosonic ones. This is also
the square of the contribution we had previously, modulo statistics.
Thus we see that the the partition function in this case is
$$Z_{n}=T^{\pm 1/2} \no$$
where the plus and minus depend on whether the fields were chosen
Bosonic or Fermionic. In the case in which the fields $f_{(n,n)}$
and $g_{(n,n)}$ are always chosen Bosonic irrespective of dimension,
the partition function is
$$\eqalign{&T^{1\over2}\qquad\hbox{ for odd $n$}\cr
  &T^{-{1\over2}}\qquad \hbox {for even $n$.}\cr}\no$$
Finally choosing a different starting field $f_{(k,k)}$ would, in a
similar fashion, yield a partition function which is $T^{\pm 1/2}$
\par
Now let us consider the case of a product manifold. We have
described what happens when we consider the gauge fixing of a
particular field $f_{(n,n)}$ on the product manifold $M_{1}\times
M_{2}$ in section 3. The further contribution that enters the
partition function is the contribution from the original action. We
take for definiteness the topological action $f_{(n,n)}df_{(n,n)}$
with $n$ even. With our gauge fixing measure \docref{measure} the
only field to be integrated over is $f_{(n,n)}^{\prime\prime}$. This
action becomes on extraction of the harmonic contributions from
$M_2$
$$f_{(n,n)}df_{(n,n)}=\widetilde f_{(n,n)}d\widetilde f_{(n,n)}
+\sum_{j=0}^{n}\sum_{a=1}^{b_{j}}f^a_{(n-j,n-j)}df^a_{(n-j,n-j)}\no$$
Integrating these fields out gives
\eqlabel{\actionx}
$$\widetilde X_{n}^{1\over 4}\prod_{j=0}^{n}
X_{j}(M_1)^{{1\over4}b_{n-j}}\no$$
Including the contribution $\widetilde X_{n}^{1/4}$ with the measure
$\widetilde\mu_{(n,n)}$ of \docref{tldbrstdet} we have
$$\widetilde Z(M_{1}\times M_{2})
=\widetilde\mu_{(n,n)}\widetilde X_{n}=1\no$$
from our analysis of degeneracies in section 2. The relevant
cancellation occurs mode by mode. In fact this does not rely on the
form of regulator used to regulate the determinants as long as it is
implemented in a uniform manner. Combining the remaining term from
the measure \docref{harmonic} with the contribution from
\docref{actionx}, together with the fact that $\widetilde
Z(M_1\times M_2)=1$, gives
$$Z(M_1\times M_2)=\prod_{k=0}^{n-1}{X_k(M_1)}^{p_k\sigma_k}
\prod_{k=0}^{n}X_k(M_1)^{b_k/4} \no$$
Noting the range of the sums concerned and using Poincar\'e duality
we see that
$$\eqalign{4\sigma_k+{(-1)}^kb_{k}&
=\sum_{j=0}^{d_2}{(-1)}^{j}b_{j}\qquad\qquad\hbox{ for each $k$}\cr
&=\chi(M_2)\qquad\qquad \hbox{ the Euler
character of $M_2$.}\cr}\no$$
The final result is that
$$Z(M_{1}\times M_{2})= Z(M_{1})^{\chi(M_{2})}\no$$
which establishes the product formula. The argument given here in
fact can also be used to demonstrate the same formula for the
combinatorial torsion.
\beginsection{Conclusion}
In this paper we have followed two primary themes, the first being a
detailed examination of the kernels of $d$ and $d^*$, the second
being the BRST quantization procedure using the gauge Fermion
construction of Batalin and Vilkovisky. We have examined both topics
on an arbitrary manifold $M$ as well as specifically  on product
manifolds. We then combined these  two pieces of work  to give a
field theoretic derivation of the formula for the Ray-Singer torsion
on product manifolds.
\par
This derivation can be seen to encompass both the Ray-Singer and
combinatorial torsions, depending on which regularisation procedure
is adopted for the field theory: Considering the procedure on a
simplicial manifold yields the combinatorial torsion, whereas
regulating the continuum theory by defining the determinants using
zeta functions yielded the analytic torsion. M\"ullers proof \ref{2}
that the torsion defined on a simplicial mesh, and invariant under
mesh refinement, is equal to the analytic torsion  provides us here
with concrete examples of field theories which exists in the
combinatorial sense and whose continuum limit is also well defined.
The equality of these two field theories under lattice subdivision
is an example of strong renormalization group invariance.
\par
We obtained several useful results which should be of use in a wider
context also. We have provided a detailed analysis of the spectral
decomposition of the eigenspaces of Laplacians on $r$-forms,
establishing the decomposition of the kernels of $d$ and $d^*$.  We
took apart this structure for product manifolds and obtained compact
expressions showing in detail how  $\ker d$ and  $\ker d^*$   are
constructed from the spectral differential form data for the factors
$M_1$ and $M_2$. This information was essential for a proper
understanding of the decomposition of the gauge Fermion on product
manifolds as well as  for the proof of the product formula for the
torsion.
\par
We constructed the gauge fixing for an $n$-form gauge
field on $M$ using the gauge Fermion method of Batalin and
Vilkovisky. By concentrating on this structure, and developing a
compact notation, we were able, very simply, to construct the
necessary measure for the quantisation of such a field. Our
procedure constructed a formalism in which the details of the
manifold $M$  play very much a subordinate r\^ole, much as is the
case in the theory of universal characteristic classes. This allowed
us to give a completely general gauge Fermion $\Psi$ which can be
viewed as a `Universal Gauge Fermion' $\Psi$---it is an infinite
dimensional upper triangular matrix, but, just as in the universal
characteristic class formalism, the finite dimensionality of the
particular $M$ chosen and the field being gauge fixed determine
which finite dimensional subset of $\Psi$ to use.
\par
We also observed that by fixing the statistics of one of the fields
at the lower apex of $\Psi$ the statistics of all other fields are
fixed. We considered both possibilities for the statistics of the
apex field---the choice, of a Fermionic apex corresponds, for
example, to gauge fixing a Bosonic connection. Our notation makes
manifest the nested structure of the gauge Fermion so that the only
modification necessary to gauge fixing an $n$-form field, rather
than an $(n-1)$-form field, is the inclusion of another row of the
gauge Fermion triangle. This is an example of how the universality
referred to in the previous paragraph operates.
\par
This nested structure suggested a new class of topological actions
of the form $f_{(n,n)}df_{(n,n)}$ which exist only for odd
dimensional manifolds; where the fields $f_{(n,n)}$ are Bosonic for
odd $n$ and Fermionic for even $n$. Each of these actions gives a
partition function which is the quarter power of the torsion.
\par
Further work that is worth considering along these lines is an
examination of the BRST procedure on a manifold where the metric is
Wick rotated to give it a Lorentzian signature. On the initial
Riemannian manifold the gauge fixing is complete, however after a
Wick rotation the gauge fixing develops singularities and no longer
appears to be complete. We have given sufficient spectral
information and a sufficiently simple method of constructing the
gauge Fermion that these and similar questions should be within easy
grasp. A further and more detailed analysis of the combinatorial
torsion would be worthy of attention. It would be nice to
examine what additional information can be gleaned from this
procedure on complex manifolds where complex torsion can be defined.
\par
Interesting examples of field theories defined discretely (in
addition to the usual lattice QCD and Regge calculus) are the
simplicial theory of Sorkin \ref{11} and the finite element theories
constructed by  Bender et al. \ref{9,10}. Much of the present work
is applicable to these latter cases. This general area deserves
further attention.

\par\vskip\baselineskip
\centerline{{\bf References}}
\vskip0.5\baselineskip

{ \par \noindent \par \hangindent \parindent \indent \hbox to\z@
{\hss \fam \bffam \tenbf 1.\kern .5em } \ignorespaces Birmingham D.,
Blau M., Rakowski M. and Thompson G., Topological field theory,
Phys. Rep., {\fam \bffam \tenbf 209}, 129--340, (1991). \par \vskip
-0.8\baselineskip \noindent } { \par \noindent \par \hangindent
\parindent \indent \hbox to\z@ {\hss \fam \bffam \tenbf 2.\kern .5em
} \ignorespaces  M{\accent "7F u}ller W., Analytic torsion and the
R-torsion of Riemannian manifolds, Adv. Math., {\fam \bffam \tenbf
28}, 233--305, (1978).\par \vskip -0.8\baselineskip \noindent } {
\par \noindent \par \hangindent \parindent \indent \hbox to\z@ {\hss
\fam \bffam \tenbf 3.\kern .5em }\ignorespaces  Ray D. B. and Singer
I. M., R-torsion and the Laplacian on Riemannian manifolds, Adv. in
Math., {\fam \bffam \tenbf 7}, 145--201, (1971).\par \vskip
-0.8\baselineskip \noindent } { \par \noindent \par \hangindent
\parindent \indent \hbox to\z@ {\hss \fam \bffam \tenbf 4.\kern .5em
}\ignorespaces  Cheeger J., Analytic torsion and the heat equation,
Ann. Math., {\fam \bffam \tenbf 109}, 259--322, (1979).\par \vskip
-0.8\baselineskip \noindent } { \par \noindent \par \hangindent
\parindent \indent \hbox to\z@ {\hss \fam \bffam \tenbf 5.\kern .5em
}\ignorespaces  Ray D. B. and Singer I. M., Analytic Torsion for
complex manifolds, Ann. Math., {\fam \bffam \tenbf 98}, 154--177,
(1973).\par \vskip -0.8\baselineskip \noindent } { \par \noindent
\par \hangindent \parindent \indent \hbox to\z@ {\hss \fam \bffam
\tenbf 6.\kern .5em }\ignorespaces  Witten E., On quantum gauge
theories in two dimensions, Commun. Math. Phys., {\fam \bffam \tenbf
141}, 153--209, (1991).\par \vskip -0.8\baselineskip \noindent } {
\par \noindent \par \hangindent \parindent \indent \hbox to\z@ {\hss
\fam \bffam \tenbf 7.\kern .5em }\ignorespaces  Batalin I. A. and
Vilkovisky G. A., Existence theorem for gauge algebras, Jour. Math.
Phys., {\fam \bffam \tenbf 26}, 172--, (1985).\par \vskip
-0.8\baselineskip \noindent } { \par \noindent \par \hangindent
\parindent \indent \hbox to\z@ {\hss \fam \bffam \tenbf 8.\kern .5em
}\ignorespaces  Batalin I. A. and Vilkovisky G. A., Quantisation of
Gauge Theories with linearly independent generators, Phys. Rev.,
{\fam \bffam \tenbf D28}, 2567--, (1983).\par \vskip
-0.8\baselineskip \noindent } { \par \noindent \par \hangindent
\parindent \indent \hbox to\z@ {\hss \fam \bffam \tenbf 9.\kern .5em
}\ignorespaces  Bender C. M., {\fam \itfam \tenit Formulation of
quantum field theory on Minkowski space-time lattice}, Paris-Meudon
Colloquium 1986, {edited by: de Vega H. and S{\accent 19 a}nchez
N.}, World Scientific, (1987). \par \vskip -0.8\baselineskip
\noindent } { \par \noindent \par \hangindent \parindent \indent
\hbox to\z@ {\hss \fam \bffam \tenbf 10.\kern .5em }\ignorespaces
Bender C. M., Cooper F., Milton K. A., Pinsky S. S. and Simmons jnr.
L. M., Discrete-time quantum mechanics. III Spin Systems, Phys.
Rev., {\fam \bffam \tenbf D35}, 3081--3091, (1987).\par \vskip
-0.8\baselineskip \noindent } { \par \noindent \par \hangindent
\parindent \indent \hbox to\z@ {\hss \fam \bffam \tenbf 11.\kern
.5em }\ignorespaces Sorkin R., The electromagnetic field on a
simplicial lattice, J. Math. Phys., {\fam \bffam \tenbf 16},
2432--2440, (1975).\par\vskip -0.8\baselineskip \noindent }

\bye
\cite{672}    
absence of a field theoretic product formula \cite{671}    
\cite{183}    
Vilkovisky: Existence theorem for gauge algebras \cite{507}
linearly independent generators \cite{697}    
Theory and Finite Elements. \cite{698}    
Theory and Finite Elements.
\bye